\documentclass[conference]{IEEEtran}
\makeatletter
\def\ps@headings{%
\def\@oddhead{\mbox{}\scriptsize\rightmark \hfil \thepage}%
\def\@evenhead{\scriptsize\thepage \hfil \leftmark\mbox{}}%
\def\@oddfoot{}%
\def\@evenfoot{}}
\makeatother
\IEEEoverridecommandlockouts
\usepackage{amsmath,amssymb}
\usepackage[mathscr]{eucal}
\usepackage{graphics,graphicx,multicol}

\begin{document}
\title{On Algebraic Traceback in Dynamic Networks}

\author{\IEEEauthorblockN{Abhik Das, Shweta Agarwal and Sriram Vishwanath}
\IEEEauthorblockA{Department of Electrical \& Computer Engineering\\ University of Texas, Austin, USA\\
Email: \{akdas, shweta.a\}@mail.utexas.edu, sriram@austin.utexas.edu}}
\maketitle

\begin{abstract}
This paper introduces the concept of {\em incremental traceback} for determining changes in the trace of a network as it evolves with time. A distributed algorithm, based on the methodology of algebraic traceback developed by Dean et al., is proposed which can completely determine a path of $d$ nodes/routers ($d\in \mathbb N$) using $O(d)$ marked packets, and subsequently determine the changes in its topology using $O(\log d)$ marked packets with high probability. The algorithm is established to be order-wise optimal i.e., no other distributed algorithm can determine changes in the path topology using lesser order of bits (i.e., marked packets). The algorithm is shown to have a computational complexity of $O(d \log d)$, which is significantly less than that of any existing non-incremental algorithm of algebraic traceback. Extensions of this algorithm to settings with node identity spoofing and network coding are also presented.
\end{abstract}


\section{Introduction}
Given the increasing number and forms of attacks on networks in recent years, developing efficient counter-measures, such as \emph{traceback}, is of significant value. In this paper, we focus on determining efficient traceback mechanisms for networks with time-varying topologies. Settings such as mobile ad-hoc networks (MANETs) are of particular interest in which we desire to use traceback towards network management and countering attacks such as denial-of-service (DoS) attack. DoS attack is arguably one of the most common forms of attack on both wire-line and wireless networks, where either a single attacker or multiple distributed attackers ``flood" a victim's link with random packets to disrupt the delivery of legitimate packets. For the Internet, IP traceback is one of the possible mechanisms for determining the source of this attack \cite{Belenky} \cite{Cheswick}. Similarly, generalized (not necessarily IP-based) traceback proves useful in determining the origin of attacks for MANETs. An important point to note is that traceback may prove useful for purposes other than countering distributed DoS attacks. For instance, it can be used for network maintenance purposes \cite{Kim}, for source/route verification and to determine location of faulty nodes in the network.

Traceback mechanisms have been traditionally studied for IP-based networks under the name of IP traceback \cite{Belenky}. The common goal in traceback literature is to perform a post-attack traceback for an IP-based network to determine the source(s) of the attack. Our paper's focus is on dynamic networks (which may or may not be IP-based) where traceback is preemptively performed to manage the network and deter possible attacks.  To this end, we desire that the traceback mechanism be efficient and be able to track changes in the traces quickly with minimal computation. In this paper, we develop an {\em incremental traceback} mechanism which, after initialization, requires a low packet and computational overhead to detect and determine changes in traces of the network.

\subsection{ Background on Traceback}
As mentioned earlier, a large body of literature on traceback focuses on IP traceback. However, regardless of the setting, good traceback mechanisms share some common properties -- they should (a)~be partially deployable in the network, (b)~result in little or no change in the router hardware, (c)~provide accurate traceback using a small number of packets, (d)~need as minimal an extent of ISP involvement as possible, (e)~perform well in presence of multiple attack sources and forms, (f)~have a low complexity mechanism for identifying attackers. These properties also serve as the evaluation metrics when comparing different traceback approaches.

The importance of the IP traceback problem has led to a large body of research in the field, resulting in the development of many interesting traceback mechanisms and methodologies to date. We briefly describe some of them:
\begin{enumerate}
\item Savage et al. \cite{Savage} proposed one of the earliest \emph{probabilistic} traceback mechanisms where routers randomly mark packets with their partial path information during the process of packet-forwarding. The main disadvantage of the scheme is the combinatorial computational complexity of the traceback process.
\item Song and Perrig \cite{Perrig} proposed an improved and authenticated packet-marking scheme with the ability to cope with multiple attacks. However, the traceback process by any workstation needs the knowledge of its current upstream router map to all attackers.
\item Bellovin et al. \cite{Bellovin} developed iTrace, a traceback scheme where routers randomly send their IP addresses in form of special packets to the source or destination IP address of the data packets. The use of special packets generate additional traffic; besides every workstation has to wait for long enough time for getting sufficient number of special packets to carry out traceback.
\item Dean et al. \cite{Algebraic} suggested a novel \emph{algebraic approach} to the IP traceback problem -- encoding the IP addresses of routers a packet passes through, into a polynomial. This allows reconstruction of the entire path in one go after getting sufficient number of packets.
\item Adler \cite{Adler} gave a detailed theoretical analysis of the traceback problem, described the tradeoffs of probabilistic packet-marking scheme and proposed a $1$-bit packet marking method to counter DoS attack.
\item Snoeren et al. \cite{Hash} proposed SPIE, a mechanism which tracks every packet through querying of the states of the upstream routers. However, this requires the routers to store a large amount of state information.
\item Thing and Lee \cite{Lee} showed that the performance of a traceback process in a wireless ad-hoc network depends on the routing protocol and network size.
\end{enumerate}

In this paper, we perform traceback in a continuous manner, with the goal of ensuring that the destination(s) in a network stay well informed of the path(s) traversed by the packets received by them. We desire that the technique used for traceback is such that each node in the network remains blind to the global network topology and the changes in it. Essentially, when a change in topology occurs, we require that the destination(s) alone detect this change and initiate an incremental traceback analysis while the remaining nodes (including the source(s)) remain oblivious to the change.

Towards the end of developing an incremental traceback mechanism with desired qualities, we use the framework of algebraic traceback as developed by Dean et al. \cite{Algebraic}. Once the algebraic traceback process is initialized using the algorithm in \cite{Algebraic}, we show that $O(\log d)$ marked packets and a traceback algorithm with a computational complexity of $O(d\log d)$ operations per execution are sufficient to {\em track the change} (node addition and deletion) in a path involving $d$ nodes ($d\in \mathbb N$). Note that, if the non-incremental algebraic traceback process were repeated each time there is a change in the path, $O(d)$ marked packets would be required to perform traceback. Next, we argue that our incremental traceback process is order-wise optimal in terms of the number of marked packets required and has a lower computational complexity compared to the conventional non-incremental traceback processes.

The rest of this paper is organized as follows. Sections \ref{sec:sysmodel} and \ref{sec:review} give the system model and a detailed review of the algebraic traceback mechanism respectively. The incremental traceback schemes based on different path encoding versions of algebraic traceback are presented in Sections \ref{sec:fullpath} and \ref{sec:randompath}. We describe the traceback procedure for systems employing network-coding in Section \ref{sec:netcod}. The numerical results are shown in Section \ref{sec:simulation} and the paper concludes with Section \ref{sec:conclude}.

\section{System Model}\label{sec:sysmodel}
We consider a network represented by a directed graph. The nodes in the graph (identifiable with routers in the network) have unique identifiers (IDs) that come from the finite field $GF(p)$, for some suitable prime number $p$. A directed edge between a pair of nodes in the graph represents an error-free channel. We assume that the transmissions across different edges do not interfere with each other in any way.

Each node can act as a source, a destination or an intermediate packet-forwarding node, depending on the communication pattern in the network. We focus our attention on one such source and destination, represented in the graph by nodes $r_1$ and $D$ respectively. The source transmit data to the destination via the path ${\mathcal P} = (r_1, r_2,\ldots, r_d, D)$. However, this path may change over the course of the transmission due to the dynamic nature of the network/graph. We want to develop an incremental algebraic traceback mechanism that enables destination $D$ to figure out this change in path $\mathcal P$.

We assume that there is the possibility of node-ID spoofing, i.e., a malicious node in path $\mathcal P$ misreporting its ID to avoid detection by destination $D$. We also limit our incremental traceback approach to track single node addition and deletion in  path $\mathcal P$. This is deliberate, as conventionally, in wireless networks, the timescale at which routes/paths change (of the order of seconds) is many orders of magnitude greater than the timescale of data transmission (of the order of milliseconds or less). Thus, any one change can be detected before additional changes occur in a path. Our algorithm and analysis framework can be naturally extended to scenarios when multiple nodes can enter or leave path $\mathcal P$. The assumption also makes the algorithm description and proofs much more intuitive and concise, and therefore we focus on this simple case.
\begin{figure}[t]
\begin{center}
\includegraphics[scale=0.30]{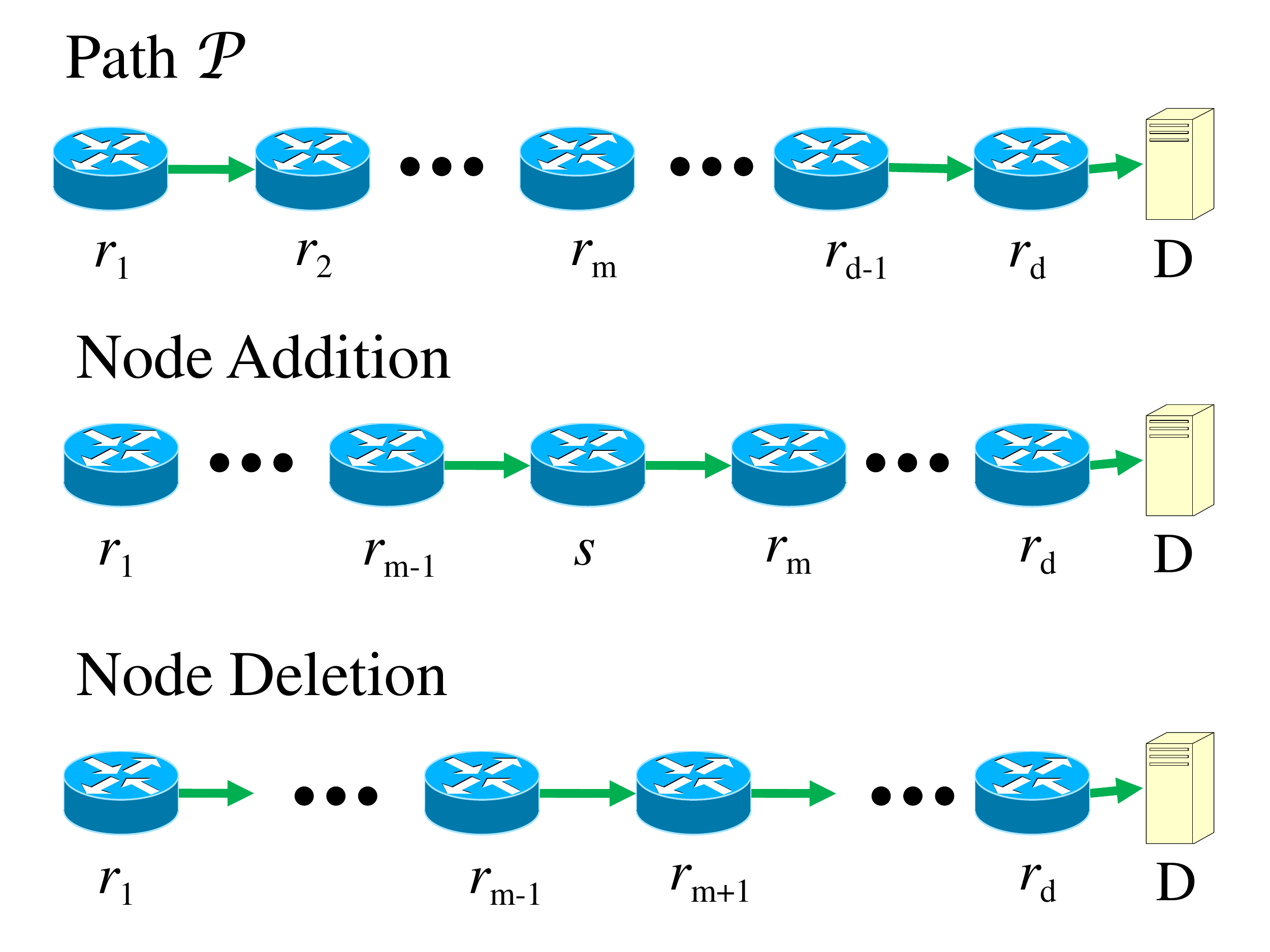}
\caption{Dynamic behavior of path $\mathcal P$}\label{fig:route}
\end{center}
\end{figure}

\section{Review: Algebraic Traceback}\label{sec:review}
In this section, we present certain relevant aspects of algebraic traceback as developed by Dean et al. \cite{Algebraic}. The idea behind this traceback scheme is that a polynomial of degree $n$ in $GF(p)$ is completely determinable using $(n+1)$ of its evaluations at distinct points in $GF(p)$. Though originally designed for IP traceback to counter DoS attack, the approach can be generalized to traceback in non-IP based networks.

\subsection{Deterministic Path Encoding}\label{sec:detpathencoding}
The deterministic path encoding scheme is used when no node-ID spoofing is suspected. The packet marking process is initiated by the first node that encounters the packet (source node, which is $r_1$ for path $\mathcal P$). We include a flag-bit field and hop-count field (with initial values $0$) in each packet in the network -- the flag-bit and hop-count values are set to $1$ when a packet is marked, otherwise the flag-bit value remains unchanged and each node following the source node just increments the hop-count by 1. In path $\mathcal P$, when node $r_1$ initiates the process of marking a packet (with some probability, say $q_1$), it encodes a value-pair $(x,y)$ into it, where $x$ is chosen randomly from $GF(p)$ and $y=r_1$. If node $r_i$ ($i=2,\ldots,d$) encounters a marked packet, it uses the values $x,y,r_i$ to update the value of $y$ as follows:
\begin{equation}
\label{update}
y\leftarrow y\cdot x + r_i.
\end{equation}
Hence, any marked packet received by destination $D$ has a value-pair of the form $(x,y(x))$ encoded in it, where
$$
y(x)=\sum_{i=0}^{d-1}r_{d-i}x^i.
$$

If destination $D$ receives $d$ value-pairs $(x_i,y(x_i)),\,\,i=1,2,\ldots,d$, where $x_i\neq x_j\,\,\forall i\neq j$, path $\mathcal P$ can be reconstructed by solving the following matrix equation:
$$
\begin{array}{rcl}
\left(\begin{array}{ccccc} 1 & x_1 & x_1^2 & \ldots & x_1^{d-1}\\ 1 & x_2 & x_2^2 & \ldots & x_2^{d-1}\\
\vdots & \vdots & \vdots & \ddots & \vdots\\ 1 & x_d & x_d^2 & \ldots & x_d^{d-1} \end{array}\right)
\left(\begin{array}{c} r_d\\ r_{d-1}\\ \vdots\\ r_1 \end{array}\right)=
\left(\begin{array}{c} y(x_1)\\ y(x_2)\\ \vdots\\ y(x_d) \end{array}\right)
\end{array}.
$$
The value of $d$ is obtained from the hop-count field of the marked packets. The resulting matrix in the equation is a full-rank Vandermonde matrix, and thus the system of equations can be solved in $O(d^2)$ operations. Thus, path $\mathcal P$ is determinable using $d$ marked packets, provided the $x$-values encoded in them are distinct. This can be ensured with high probability by making the source $r_1$ keep a record of the $x$-values it has used while marking packets, thereby avoiding re-use of the $x$-values until the marking of at least $p$ packets. Therefore, choosing a large enough $p$ can ensure that $O(d)$ marked packets are sufficient for retrieval of path $\mathcal P$.

\subsection{Randomized Path Encoding}\label{sec:randpathencoding}
The deterministic path encoding scheme may be infeasible if node-ID spoofing is possible and/or the first node to receive a packet is unsure if it is indeed the source node (for example, if $r_1$ does not know it is the source node in path $\mathcal P$). Then we require a probabilistic traceback mechanism to address this situation. For path $\mathcal P$, node $r_1$ initiates marking of the packet as before (with probability $q_1$), but now each intermediate node $r_i$ ($i=2,\ldots,d$) clears an existing marking, if any, and re-marks a packet with probability $q_i$. Else, with probability $(1-q_i)$, each node $r_i$ just follows the update mechanism as given by (\ref{update}). The following pseudo-code summarizes this procedure:\\
\emph{Marking scheme at node} $r_i$:\\
\hspace*{1 cm} for each packet $w$\\
\hspace*{1 cm} with probability $q_i$\\
\hspace*{2 cm} $x = random$;\\
\hspace*{2 cm} $y = r_i$;\\
\hspace*{2 cm} $flagbit = 1$;\\
\hspace*{2 cm} $hopcount = 1$;\\
\hspace*{1 cm} otherwise if $flagbit = 1$\\
\hspace*{2 cm} $y \leftarrow y \cdot x + r_i$;\\
\hspace*{2 cm} $hopcount \leftarrow hopcount + 1$;

We assign non-trivial values to the marking probabilities $q_i,\,i=1,2,\ldots,d$ such that the traceback process remains accurate while not requiring a very large overhead. For example, \cite{Algebraic} examines the case when $q_i=q\in (0,1)\,\,\forall i$. Then, apart from marked packets with value-pairs corresponding to path $\mathcal P$, there are marked packets with value-pairs corresponding to sub-paths ${\mathcal P}_i=(r_{i+1},r_{i+2},\ldots,r_d),\,i=1,2,\dots,d-1$, as well. A marked packet received by destination $D$ has a value-pair of the form $(x,y(x))$ where
$$
y(x)=\sum_{i=0}^{k}r_{d-i}x^i,\,\,k=0,1,\ldots,d-1.
$$
These marked packets can be segregated, in terms of the sub-paths their value-pairs correspond to, on the basis of their hop-count values\footnote{For simplicity, we assume that the hop-count field is not attacked. If this field is attackable, then alternate mechanisms for path reconstruction exist such as the Guruswami-Sudan algorithm based mechanism presented in \cite{Algebraic}.}, as a hop count of $i$ $(<d)$ implies that the value-pair is for ${\mathcal P}_{d-i}$ and, consequently, a hop-count of $d$ implies that the value-pair is for path $\mathcal P$. Using this, the sub-paths and therefore, the entire path $\mathcal P$ can be reconstructed after getting sufficient number of marked packets, in a manner similar to deterministic path encoding. The $x$-values across nodes can be maintained as distinct values (to ensure invertibility of the resulting matrix at the destination) by requiring that the nodes with non-zero marking probabilities keep a track of the $x$-values they use while marking packets and only reuse values when all elements in $GF(p)$ have been exhausted.

Suppose $f_i,\,i=1,2,\ldots,d$ be defined as the fraction of packets marked by node $r_i$ and received by destination $D$, then $f_i$ can be expressed in terms of $q_i,\,i=1,2,\ldots,d$, as
$$
f_i = \left\{\begin{array}{cl} q_i\prod_{j=i+1}^d(1-q_j) & \textrm{if} \,i\neq d\\
q_d & \textrm{if} \,i=d
\end{array} \right.
$$
with the fraction of unmarked packets given by $f_0=1-(\sum_{i=1}^d f_i)=\prod_{i=1}^d(1-q_i)$. This makes the fraction of marked packets coming from source $r_1$ to be $\frac{f_1}{1-f_0}$, i.e., one out of $\lceil\frac{1-f_0}{f_1}\rceil$ marked packets is from node $r_1$ on an average. Since $d$ marked packets from node $r_1$ with distinct $x$-values are needed for determining path $\mathcal P$, an average of $d\lceil\frac{1-f_0}{f_1}\rceil$ marked packets needs to be received by destination $D$ to ensure that $d$ packets among them have value-pairs corresponding to path $\mathcal P$.

If $q_i=q \,\,\forall i$, we have $f_0=(1-q)^d$ and $f_1=q(1-q)^{d-1}$, which gives the average number of marked packets as
$$
d\left\lceil\frac{1-f_0}{f_1}\right\rceil=d\left\lceil\frac{1-(1-q)^d}{q(1-q)^{d-1}}\right\rceil
=d\left\lceil\sum_{i=0}^{d-1}\frac{1}{(1-q)^i}\right\rceil.
$$
As $q\rightarrow 0$, the above quantity goes to $d^2$. Hence, if $q$ is chosen reasonably small, an average of $O(d^2)$ marked packets are sufficient for determining path $\mathcal P$. But $f_0$ is large for small $q$, which is inefficient as then destination $D$ has to wait for a longer time to receive sufficient number of marked packets for performing traceback. Thus, there is a tradeoff in the value of $q$. Even for the general case of marking probabilities, $d\lceil\frac{1-f_0}{f_1}\rceil$ becomes smaller as $f_0$ and $f_1$ become large. But $f_0$ cannot be very large, causing a tradeoff. Regardless of this tradeoff, an average of $O\left(d(\frac{1-f_0}{f_1})\right)$ marked packets is necessary.

\section{Inc. Traceback: Deterministic Path Encoding}\label{sec:fullpath}
In this section, we present an incremental traceback approach, based on the methodology of deterministic path encoding. We adopt the same encoding/marking procedure i.e., the source node initiates the packet marking process. As discussed earlier, path $\mathcal P$ can be ascertained using $O(d)$ marked packets with a computational complexity of $O(d^2)$. Our interest is in the case when this initial process has occurred, and then path ${\mathcal P}$ changes due to node addition or deletion. A conventional traceback mechanism would repeat the traceback procedure again, i.e., destination $D$ would wait until it receives $O(d)$ marked packets again, reconstruct the modified path and then determine where the change has occurred. This scheme proves to be inefficient -- the number of marked packets and computational load incurred remains the same. The proposed incremental traceback method makes use of the fact that path $\mathcal P$ is known to destination $D$ (due to an initial traceback process) to determine the change using $O(\log d)$ marked packets with a computational complexity of $O(d\log d)$.

The change in topology of path $\mathcal P$ involves either addition or deletion of a single node, which can be detected using the hop-count value of a marked packet -- it changes from $d$ to $(d+1)$ for node addition and to $(d-1)$ for node deletion. We examine these two cases separately.

\subsection{Node Addition}
Note again that the encoding process remains the same as before (as in Section \ref{sec:detpathencoding}). In incremental traceback, all that changes is the decoding algorithm at the destination $D$. Suppose a node with ID $s$ gets added to path $\mathcal P$ in the $m$th position, $1\leq m\leq d+1$ ($1$st position refers to the position before node $r_1$ and $(d+1)$th position refers to the position after node $r_d$). Then the new packets have value-pairs of the form $(x,z(x))$ encoded in them, where
\begin{equation}\label{eqn:nodeaddition}
z(x) = a_m(x) + x^{d-m+1}(s + x b_m(x)).
\end{equation}
$a_k(x)$ and $b_k(x)$ are polynomials given by
\begin{equation}\label{eqn:polya}
a_k(x) = \left\{\begin{array}{cl} r_d+r_{d-1}x+\ldots+r_kx^{d-k} & \textrm{if} \,k\neq d+1\\
0 & \textrm{if} \,k=d+1
\end{array} \right.
\end{equation}
\begin{equation}\label{eqn:polyb}
b_k(x) = \left\{\begin{array}{cl} r_{k-1}+r_{k-2}x+\ldots+r_1x^{k-2} & \textrm{if} \,k\neq 1\\
0 & \textrm{if} \,k=1
\end{array} \right.
\end{equation}
for $k=1,2,\ldots,d+1$. These polynomials are known to destination $D$ from the usual traceback performed previously, which gives $r_1,r_2,\ldots,r_d$. The polynomials also satisfy
$$
y(x) = \sum_{i=0}^{d-1}r_{d-i}x^i = a_k(x) + x^{d-k+1}b_k(x)
$$
$\forall k$, where $y(x)$ refers to the $y$-value of the marked packet received by destination $D$ prior to addition of node $s$.

Suppose $(x_i,z_i),\,i=1,2,\ldots,l$, are the value-pairs encoded in $l$ marked packets received after the addition of $s$ in path $\mathcal P$. We consider the following set of equations:
\begin{equation}\label{eqn:eqnsetnodeaddition}
z_j=a_k(x_j)+x_j^{d-k+1}(s+x_jb_k(x_j)),\,\,1\leq j\leq l.
\end{equation}
From (\ref{eqn:nodeaddition}), the set of equations is consistent for $k=m$. For $k\neq m$, the set of equations is not consistent with high probability (this is established by Theorem 1 below). We make use of this property to design an incremental traceback algorithm for destination $D$ as follows:\\
\emph{Algorithm I}
\begin{enumerate}
\item Construct a $(d+1)\times l$ matrix ${\hat S}=[{\hat s}_{kj}]$ where
$$
{\hat s}_{kj} = \frac{z_j - a_k(x_j)}{x_j^{d-k+1}} - x_j b_k(x_j).
$$
\item If there exists a unique row in ${\hat S}$ with equal elements, say the $\hat m$th row, declare that the new node is in $\hat m$th position with ID $\hat{s} = {\hat s}_{\hat{m}j},\,1\leq j\leq l$.
\item If there exists more than one row in $\hat S$ with equal elements, declare that an error has occurred. Wait for more value-pairs to arrive through marked packets, say $(x_i,z_i),\,i=l+1,\ldots,l+\epsilon$, where $\epsilon$ is an integer of smaller order compared to $l$. Repeat the algorithm using the value-pairs $(x_i,z_i),\,i=\epsilon+1,\ldots,l+\epsilon$. Theorem 1 below shows that the algorithm terminates with high probability while obtaining the correct node ID.
\end{enumerate}

\noindent {\em Theorem 1:} A newly added node in path $\mathcal P$ can be identified by destination $D$ using $l=O(\log d)$ marked packets and \emph{Algorithm I}, with a computational complexity of $O(d\log d)$.
\\
{\em Proof:} From (\ref{eqn:eqnsetnodeaddition}), it is clear that all elements of the $m$th row of $\hat S$ will be equal. If this is the only such row, we have the correct new node position and ID $s = {\hat s}_{mj},\,1\leq j\leq l$. An error occurs if there exists another row $i \ne m$ such that all elements of the $i$th row are equal as well. To determine the probability of this happening, we note that $x_j$ is chosen uniformly over $GF(p)$. This makes ${\hat s}_{kj}$ uniform for any $k\neq m$, since each ${\hat s}_{kj}$ is purely a function of $x_j$. So, ${\hat s}_{ij},\,j=1,2,\ldots,l$ is an i.i.d. uniform random process. This gives
$$
Pr({\hat s}_{ij} = {\hat s}_{ij'}) = \frac{1}{p} = 2^{-\log_2 p}
$$
for any $1\leq j,j'\leq l$ and $j\neq j'$. Let $E_i$ be the event that all elements of the $i$th row of ${\hat S}$ are same. Then we have $Pr(E_i) = 2^{-l \log_2 p}$ for $i\neq m$, since there are $l$ elements in each row. The probability of error is
$$
P_e = Pr(\cup_{i \ne m} E_i) \leq dPr(E_i)= 2^{\log_2 d-l \log_2 p}
$$
where the inequality above is due to the union bound. $P_e$ can be made arbitrarily small if $\log_2 d-l \log_2 p$ can be made as negative as possible. If we require that $l > \frac{\log_2 d}{\log_2 p}$, then this can be satisfied. Thus, we choose $l= \lceil\frac{\log_2 d}{\log_2 p}+\delta \rceil$, where $\delta \in \mathbb{N}$ is a small constant. Then $P_e$ gets upper-bounded as
$$
P_e\leq 2^{\log_2{d}-l\log_2{p}}=\frac{1}{p^{\delta}}2^{\log_2{d}-\log_2{p}\lceil\frac{\log_2 d}{\log_2 p}\rceil}\leq \frac{1}{p^{\delta}}
$$
where the second inequality follows from the fact that $a-b\lceil\frac{a}{b}\rceil \leq 0\,\,\forall a,b\in \mathbb{R},\,b\neq 0$. By choosing a large enough value for $p$, $P_e$ can be bounded above by any arbitrary small positive value. In other words, $l=O(\log d)$ is sufficient for determining the newly added node correctly.

Since the algorithm relies on the computation of $\hat{S}$ which has $(d+1)l$ entries, we get a complexity of $O(d\log d)$ (since $l=O(\log d)$). This completes our proof.

\subsection{Node Deletion}
Suppose node $r_m$ ($1\leq m\leq d$) gets deleted from path $\mathcal P$, leaving behind $d-1$ nodes. Then the new marked packets carry value-pairs of the form $(x,w(x))$, where
\begin{equation}\label{eqn:nodedeletion}
w(x) = a_m(x) - x^{d-m}(r_m - b_m(x)).
\end{equation}
$a_k(x)$ and $b_k(x)$ are polynomials as defined in (\ref{eqn:polya}) and (\ref{eqn:polyb}).

Suppose $(x_i,w_i),\,i=1,2,\ldots,l$ be the received value-pairs from $l$ marked packets received after deletion of node $r_m$. We consider the following set of equations:
\begin{equation}\label{eqn:eqnsetnodedeletion}
w_j=w(x_j)=a_k(x_j)-x_j^{d-k}(r_k-b_k(x_j)),\,\,1\leq j\leq l.
\end{equation}
From (\ref{eqn:nodedeletion}), the set of equations is consistent for $k=m$. For $k\neq m$, the set of equations is not consistent with high probability (proved in Theorem 2). We make use this property to design an incremental traceback algorithm for destination $D$, for the case of node deletion, as follows:\\
\emph{Algorithm II}
\begin{enumerate}
\item Construct a $d\times l$ matrix ${\hat R}=[{\hat r}_{kj}]$ where
$$
{\hat r}_{kj} = b_k(x)-\frac{w_j - a_k(x_j)}{x_j^{d-k}}.
$$
\item If there exists a unique row in ${\hat R}$ with equal elements, say the $\hat m$th row, declare that the deleted node was in $\hat m$th position with ID $\hat{r} = {\hat r}_{\hat{m}j},\,1\leq j\leq l$.
\item If there exists more than one row in $\hat R$ with equal elements, declare that an error has occurred. Wait to receive more value-pairs through marked packets, say $(x_i,w_i),\,i=l+1,\ldots,l+\epsilon$, where $\epsilon$ is an integer of smaller order compared to $l$. Repeat the algorithm using the value-pairs $(x_i,z_i),\,i=\epsilon+1,\ldots,l+\epsilon$. Theorem 2 below shows that the algorithm terminates with high probability while obtaining the correct node ID.
\end{enumerate}

\noindent {\em Theorem 2:} A deleted node in path $\mathcal P$ can be identified by destination $D$ using $l=O(\log d)$ marked packets and \emph{Algorithm II}, with a computational complexity of $O(d\log d)$.\\
{\em Proof:} From (\ref{eqn:eqnsetnodedeletion}), all elements of the $m$th row of $\hat R$ will be equal. If this is the only such row, we have the correct deleted node ID $r_m = {\hat r}_{mj},\,1\leq j\leq l$. An error occurs if there exists another row $i \ne m$ such that all elements of the $i$th row are equal as well. Using the same argument as in the proof of \emph{Theorem 1}, we get ${\hat r}_{ij},\,j=1,2,\ldots,l$ to be an i.i.d. uniform random process. This gives
$$
Pr({\hat r}_{ij} = {\hat r}_{ij'}) = \frac{1}{p} = 2^{-\log_2 p}
$$
for $1\leq j,j'\leq l$ and $j\neq j'$. Let $E_i$ be the event that all elements of the $i$th row of ${\hat R}$ are same. Then $Pr(E_i) = 2^{-l \log_2 p}$ for $i\neq m$, and the probability of error is
$$
P_e = Pr(\cup_{i \ne m} E_i) \leq (d-1)Pr(E_i) < 2^{\log_2 d-l \log_2 p}
$$
where the inequality is again due to union bound. Since the upper-bound of $P_e$ is same as that for the case of node addition, using the same approach as in the proof of \emph{Theorem 1}, we conclude that $P_e$ can be bounded above by any arbitrary small positive value and $l=O(\log d)$ is sufficient for determining the deleted node's location and ID with high probability.
Since the algorithm makes use of $\hat{R}$, which has $dl$ entries, this results in a computational complexity of $O(d\log d)$ ($l=O(\log d)$). This completes our proof.

Thus, be it node addition or deletion, $O(\log d)$ marked packets are always sufficient for destination $D$ to determine the change in path $\mathcal P$ accurately. Before we proceed to randomized traceback algorithms, a quick note on the order-wise optimality of \emph{Algorithms I} and \emph{II}. Note that, from principles of information theory \cite{Cover}, it is well known that the entropy of a uniform source with an alphabet of size $k$ is $\log_2 k$ bits. Thus, even if a centralized mechanism existed to communicate the location of the node being inserted/deleted, it would require $O(\log_2 d)$ bits to do so, as there are $d$ equally likely places for the change. Our distributed mechanism uses $\lceil\frac{\log_2 d}{\log_2 p} + \delta\rceil$ packets or approximately $2(\log_2 d + \delta \log_2 p)$ bits. Thus, in terms of the order of growth of network overhead in $d$, the incremental traceback mechanism is order-wise optimal.

\section{Inc. Traceback: Randomized Path Encoding}\label{sec:randompath}
In this section, we present an incremental traceback approach, useful when node-ID spoofing is suspected, utilizing the randomized path encoding framework. In this setup, each packet decides to clear any existing marks and re-initiate the marking process with some probability $q_i$. As multiple nodes on path $\mathcal P$ now act as source nodes, we receive different (sub) polynomial evaluations across time. The marked packets carry value-pairs corresponding to both sub-paths $\mathcal P_i,\,i=1,2,\ldots,d-1$ and of the entire path $\mathcal P$. As described in Section \ref{sec:randpathencoding}, path $\mathcal P$ can be initially determined using an average of $O\left(d(\frac{1-f_0}{f_1})\right)$ marked packets with a computational complexity of at least $O(d^2)$. Once path $\mathcal P$ is known to the destination, we show that it possible to track its changes using lesser number of marked packets with lower complexity.

Due to the random nature of packet-marking, one cannot immediately ascertain if node addition or node deletion has occurred from the hop-count value of the marked packets. So, we need to consider both the possibilities jointly in our analysis. If a node with ID $s$ gets added to path $\mathcal P$, the value-pair of a new marked packet has information about $s$ encoded in it, provided it has traversed a sub-path containing node $s$. Similarly, if node $r_m$ is removed from path $\mathcal P$, only those marked packets that traverse sub-paths that contained node $r_m$ prior to its deletion can provide information about $r_m$.

Note that the number of marked packets required to detect a change (addition or deletion) in path $\mathcal P$ is highest when the change occurs in the first position of the path i.e., either when node $r_1$ gets deleted or a new node gets added before it. In such a situation, the marked packets that are useful in tracking this change are ones that are marked by the first node and by no other node along the new path, which we call ${\mathcal P}'$. Let $f_i'$ denote the fraction of packets received by the destination and marked by the $i$th node in path ${\mathcal P}'$. Then, the fraction of marked packets originating at the first node along path path ${\mathcal P}'$ is $\frac{f_1'}{1-f_0'}$ where $f_0'=1-(\sum_{i\geq 1}f_i')$ is the fraction of unmarked packets. This implies that, from an average of $l\lceil\frac{1-f_0'}{f_1'}\rceil$ new marked packets received by the destination after a change (addition or deletion in the path), $l$ marked packets with the highest hop-counts are likely to come from the node in the first position on path $\mathcal P'$. In the following sections, we show that $l=O(\log d)$ is sufficient to determine the ID, position and nature of the change in the path ${\mathcal P}$, given that the destination already has knowledge of the path ${\mathcal P}$.

Let us start with the assumption that a new node $s$ gets added at the $m$th position in path $\mathcal P$ ($1\leq m\leq d+1$), Now, a marked packet with hop-count $h$, where $d-m+2\leq h\leq d+1$, contains information that includes the ID $s$. Therefore, the value-pair for this packet can be rewritten as
\begin{equation}\label{eq:randomnodeaddition}
z(x)=a_m(x)+x^{d-m+1}(s+xb_{m,h}(x)).
\end{equation}
$a_k(x)$ is defined as in (\ref{eqn:polya}) and $b_{k,h}(x)$ is defined as
$$
b_{k,h}(x) = r_{k-1}+r_{k-2}x+\ldots+r_{d-h+2}x^{k-d+h-3}
$$
for $k=d-h+2,\ldots,d+1$ and $b_{k,h}(x)=0$ for $k=d-h+2$. Similarly, if node $r_m$ ($1\leq m\leq d$) is deleted from path $\mathcal P$, then a marked packet with hop-count $h$, where $d-m+1\leq h\leq d-1$ contains value-pair $(x,w(x))$ such that
\begin{equation}\label{eq:randomnodedeletion}
w(x)=a_m(x)-x^{d-m}(r_m-b_{m,h+2}(x)).
\end{equation}

Depending on whether a node gets added or deleted in path $\mathcal P$, path ${\mathcal P}'$ has $d+1$ or $d-1$ nodes respectively. Note that, if there is no change in $\mathcal P$, we have ${\mathcal P}'={\mathcal P}$. So, $f_0'$ and $f_1'$ can take three possible values, one is the unchanged $f_0$ and $f_1$, the other two values result from a change in $\mathcal P$ (node addition and node deletion). Let $F_0$ and $F_1$ denote those values of $f_0'$ and $f_1'$ that maximizes $\frac{1-f_0'}{f_1'}$ among these three choices. Suppose $(x_i,z_i),\,i=1,2,\ldots,l$ are the value-pairs of the marked packets with the highest hop-count values, say $h_i,\,i=1,2,\ldots,l$, among $l\lceil\frac{1-F_0}{F_1}\rceil$ marked packets received by the destination. Then, by an expected/average value argument, these $l$ packets are marked by nodes close to node $r_1$ and possess information about the change in path $\mathcal P$. If $h_i=d+1$ for some $i$, it means there has been node addition but if $h_i\leq d\,\,\forall i$, we cannot conclude anything and have to consider both the possibilities of node addition and node deletion. We propose the following incremental traceback algorithm for destination $D$ to determine change in path $\mathcal P$:\\
{\em Algorithm III}
\begin{enumerate}
\item  Construct a $(d+1)\times l$ matrix ${\hat S}=[\hat{s}_{kj}]$ where
$$
{\hat s}_{kj} = \frac{z_j - a_k(x_j)}{x_j^{d-k+1}} - x_j b_{k,h_j}(x_j)
$$
for $k\geq d-h_j+2$ and ${\hat s}_{kj}=0$ otherwise.
\item If there exists a unique row in ${\hat S}$, say the $\hat m$th row, such that all non-zero elements (there should be atleast two non-zero elements) of the row are equal, declare that there is a new node added in $\hat m$th position with ID $\hat{s}$ equal to the non-zero element value.
\item If there exists more than one row in $\hat S$ with equal non-zero elements, declare that an error has occurred. Wait to get more value-pairs with high hop-count values through marked packets. Repeat (i), (ii) using these and some of the earlier value-pairs ($l$ value-pairs in all).
\item If there exists no row in $\hat S$ with equal non-zero elements, construct a $d\times l$ matrix ${\hat R}=[\hat{r}_{kj}]$ where
$$
{\hat r}_{kj} = b_{k,h_j+2}(x)-\frac{z_j - a_k(x_j)}{x_j^{d-k}}
$$
for $k\geq d-h_j+1$ and ${\hat r}_{kj}=0$ otherwise.
\item If there exists a unique row in ${\hat R}$, say the $\hat m$th row, such that all non-zero elements of the row are equal, declare that the node in $\hat m$th position has been deleted with ID equal to the non-zero element value.
\item If there exists more than one row in $\hat R$ with equal non-zero elements, declare that an error has occurred. Wait to get more value-pairs with high hop-count values through marked packets. Repeat (iv), (v) using these and some of the earlier value-pairs ($l$ value-pairs in all).
\item If there exists no row in $\hat R$ with equal non-zero elements, declare that there has been no change in $\mathcal P$.
\end{enumerate}

\noindent {\em Theorem 3:} Any change in path $\mathcal P$ can be identified by destination $D$ using $l=O(\log d)$ marked packets, containing information about the change encoded in them, and {\em Algorithm III} with a computational complexity of $O(d\log d)$.\\
{\em Proof:} The cases  of node addition and node deletion cannot return positive results simultaneously i.e., both $\hat S$ and $\hat R$ cannot have unique rows with their non-zero elements equal. Since the value-pairs from the $l$ marked packets are assumed to possess information about the change in $\mathcal P$, equality of all the elements, not the non-zero elements alone, of some row of $\hat R$ or $\hat S$ would confirm the change (from (\ref{eq:randomnodeaddition}) and (\ref{eq:randomnodedeletion})). So, we need to show that, for  node addition (node deletion), the existence of more than one row in $\hat S$ ($\hat R$)  with equal elements is highly improbable for $l=O(\log d)$. Note that this is exactly what we have already established as part of the proofs of \emph{Theorems 1} and \emph{2}. Also, {\em Algorithm III} requires evaluating both $\hat R$ and $\hat S$ in the worst-case situation, each of which has a computational complexity of $O(d\log d)$. This gives an overall complexity of $O(d\log d)$. This completes our proof.

Thus, $l=O(\log d)$ marked packets, with the information of path change encoded in them, and an average of $O\left((\log d)(\frac{1-F_0}{F_1})\right)$ marked packets in general, are sufficient to determine the correct change in topology of $\mathcal P$.

\subsection{Reducing the requirement on number of marked packets}\label{subsec:marked}
In this section, we develop two schemes that enable us to reduce the average order of marked packets needed to perform probabilistic traceback. If $q_i=q \,\,\forall i$, then $f_0=(1-q)^d$, $f_1=q(1-q)^{d-1}$ and
\begin{equation}\label{eq:increase}
\frac{1-f_0}{f_1}=\frac{1-(1-q)^d}{q(1-q)^{d-1}}=\sum_{i=0}^{d-1}\frac{1}{(1-q)^i} .
\end{equation}
Since the quantity in (\ref{eq:increase}) increases with $d$, we have $\frac{1-F_0}{F_1}=\sum_{i=0}^{d}\frac{1}{(1-q)^i}$, which approaches $(d+1)$ as $q\rightarrow 0$. So, if $q$ is chosen arbitrarily small, an average of $O(d\log d)$ marked packets are sufficient for determining any change in $\mathcal P$. However, a small $q$ implies a larger value for $f_0$, and thus there is a tradeoff between the two parameters.

To reduce the average number of marked packets, we must attempt to make each of the $f_i$ values comparable to one another for this. One way this can be done is through requiring that the marking probability of a packet be dependent on the hop-count, i.e., higher the hop-count value of a packet, lesser is the probability that a node marks it. So, we have $q_i=q(h)$ where $h$ is the hop-count of a packet and $q:\mathbb{N}\rightarrow [0,1)$ is a non-increasing function in $h$. This gives $f_1 = q(1)\prod_{i=2}^d (1-q(i))$ and $f_0 = \prod_{i=1}^d (1-q(i))$ for $\mathcal P$. Next, we present two packet marking schemes with the aim of reducing the average number of marked packets needed for incremental probabilistic traceback.

\subsubsection{Scheme 1}
We consider a constant $h_0 \in \mathbb{N}$ and the following marking-probability function:
$$
q(h) = \left\{\begin{array}{cl} q \in (0,1) & \textrm{if} \, 1\leq h\leq h_0\\
0 & \textrm{otherwise}
\end{array} \right.
$$
This gives $f_1=q(1-q)^{h_0-1}$, $f_0=(1-q)^{h_0}$ and
\begin{equation}\label{quan1}
\frac{1-f_0}{f_1}=\frac{1-(1-q)^{h_0}}{q(1-q)^{h_0-1}}=\sum_{i=0}^{h_0-1}\frac{1}{(1-q)^i}
\end{equation}
for $d\geq h_0$.  As $q\rightarrow 0$, the quantity in (\ref{quan1}) goes to $h_0$. So, the average order of marked packets becomes $O(h_0\log d)=O(\log d)$ for $d\geq h_0$. Next, we substitute $q=\frac{1}{h_0}$ and get:
\begin{equation}\label{fraction1}
\frac{1-F_0}{F_1}=h_0\left[\frac{1-\left(1-\frac{1}{h_0}\right)^{h_0}}{\left(1-\frac{1}{h_0}\right)^{h_0-1}}\right]
\end{equation}
for $d\geq h_0$. As $h_0$ increases, the numerator and denominator of (\ref{fraction1}) approach $1-\frac{1}{e}$ and $\frac{1}{e}$ respectively. This makes $\frac{1-F_0}{F_1} \approx (e-1)h_0$. Also $F_0 \approx \frac{1}{e}$ i.e., about $37\%$ of the packets remain unmarked in this scheme.
\subsubsection{Scheme 2}
We consider the same constant $h_0$ and the following marking-probability function:
$$
q(h) = \left\{\begin{array}{cl} {\alpha}^h  &  \, \alpha \in (0,1),\,1\leq h\leq h_0\\
0 & \textrm{otherwise}
\end{array} \right.
$$
This gives $f_1=\alpha\prod_{i=2}^{h_0}(1-{\alpha}^i)$, $f_0=\prod_{i=1}^{h_0}(1-{\alpha}^i)$ and
\begin{equation}
\label{eq:second}
\frac{1-f_0}{f_1}=1+\frac{1}{\alpha}\left[\frac{1}{\prod_{i=2}^{h_0}(1-{\alpha}^i)}-1\right]
\end{equation}
for $d\geq h_0$. As $\alpha \rightarrow 0$, the ratio in (\ref{eq:second}) goes to $1$ and the average number of marked packets in the system is $O(\log d)$ for $d\geq h_0$. Note that there is a tradeoff in the choice of $\alpha$ - if it is small, then the fraction of unmarked packets is large. For $\alpha\in (0,\frac{1}{2}]$ and $h_0\geq 3$, we get
\begin{equation}
\label{eq:ratio2}
\frac{1-F_0}{F_1} \approx 1+\frac{\alpha}{(1-{\alpha})(1-{\alpha}^3)}.
\end{equation}
As $\alpha$ varies from very small to $\frac{1}{2}$, the quantity in (\ref{eq:ratio2}) varies from $1$ to $2\frac{1}{7}$ and the fraction of unmarked packets changes from close to $100\%$ to around $30\%$.

Thus, with an intelligent choice of marking probabilities, we can reduce the overall network overhead incurred.

\section{Traceback for Network Coding}
\label{sec:netcod}

In the previous sections, we have focused only on a single path $\mathcal P$ with source node $r_1$ and destination $D$. However, a general graph can have a multicast set-up with a source communicating to more than one destinations. In such a situation, adopting schemes such as network coding can help increase the set of rates achievable by the sources in the network. We use the algebraic traceback framework in this paper to develop a non-incremental (and incremental) mechanism of performing traceback in network coded systems.

To better motivate our traceback mechanism, we start with a simple unicast communication setup without network coding. Here, one source communicates with only one destination through a number of paths (Sections I through V have considered the case where there is just one path that is being traced). Note that, for unicast communication, network coding is not required and the Ford-Fulkerson algorithm \cite{FF} gives us routes that achieve capacity. For a network with unit capacity links and a mincut of $R$, Ford-Fulkerson returns $R$ distinct paths from source to destination. We labels these paths as ${\mathcal L_i},\,i=1,2,\ldots,R$ and the goal of traceback is to determine the identities of the nodes involved along each path at the destination. Note that, if the network mincut is $R$, the destination receives at least $R$ packets at every time instant. Here, we assume that the destination can determine which path $\mathcal L_i$ a particular packet traversed. For example, if each path were along a different OFDM sub-channel (in a MANET), then our assumption implies that the destination can identify the sub-channel through which each packet is received. Now, both the non-incremental and incremental traceback schemes described in Sections \ref{sec:review}, \ref{sec:fullpath} and \ref{sec:randompath} can be performed individually on each of the $\mathcal L_i$'s separately, and nodes along all $R$ paths between source and destination can be identified.

\begin{figure}
\begin{center}
\includegraphics[scale=0.30]{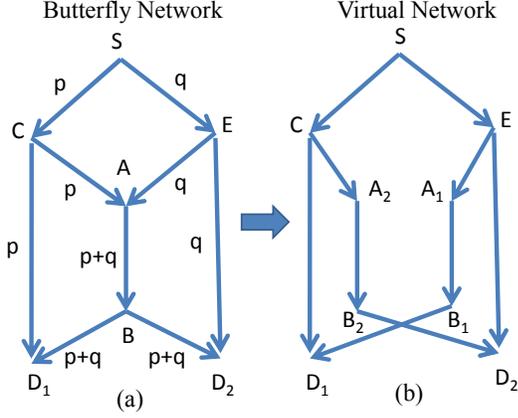}
\end{center}
\caption{The butterfly network and its equivalent virtual network}\label{fig:butterfly}
\end{figure}

Next, consider a multicast setup where in-network coding is used. In other words, there are nodes which generate (random) linear combinations of packets which they receive, and forward these combinations. We desire to develop a marking scheme that will enable us to trace the path taken by the source packet even after being linearly combined at the intermediate node with other packets.
To make our strategy concrete, we take the well-known `butterfly' network as an example for our graph (Figure \ref{fig:butterfly}). Note that our traceback procedure is in no way limited to this butterfly network and can be generalized to other multicast networks employing network coding.

In Figure 2,  $S$ is the source node and $D_1$ and $D_2$ are the destination nodes. The paths which are used by packets originating from $S$ to $D_1$ are $SCD_1$, $SEABD_1$ and from $S$ to $D_2$ are $SED_2$, $SCABD_2$ for communicating with $D_2$. Note that the min-cut for this network is 2 bits, a rate of $2$ for both $(S,D_1)$ and $(S,D_2)$ is achievable using network coding. To develop our traceback procedure, consider the virtual network in Figure \ref{fig:butterfly}-b where nodes $A$ and $B$ get split into two new node-pairs $(A_1,A_2)$ and $(B_1,B_2)$. In this virtual network, the same rate of $2$ is achievable for both $(S,D_1)$ and $(S,D_2)$ without network coding. Moreover, Ford-Fulkerson (routing) is sufficient to achieve capacity, and a traditional algebraic packet marking scheme is sufficient to perform traceback. Thus, for the original network in Figure \ref{fig:butterfly}-a, we desire to ``mimic" the virtual network in Figure \ref{fig:butterfly}-b. Say $(x_1,y_1)$ and $(x_2,y_2)$ are the value-pairs received by $A$ from $C$ and $E$ respectively, Then $A$ chooses one of the value-pairs with some probability, say $(x_i,y_i)$, and updates it using its own ID $a$, to get $(x_i,y_i')$, where $y_i'\leftarrow y_i\cdot x_i+a$. To ensure that the same path is not chosen every time, node $A$ may change the probability of selection in every time-slot. When the chosen value-pair is received by the other nodes, the same policy as traditional marking is followed. In this way a destination can determine the paths to all the sources. For example, destination $D_1$ can determine the paths $SCD_1$, $SEABD_1$ and $SCABD_1$. Thus, every destination can recreate the network subgraph corresponding to packets it observes.

\subsection{Faulty/Malicious Nodes in Network-Coded Systems}
As described above, a destination in a network-coded system traces a subgraph instead of a path traversed by a packet. Here, we describe an approach to identify a malicious/faulty node in such a network. We restrict our attention to the case in which a single node in the network is faulty or malicious; this approach can be extended to the more general case.

The broad idea is that routing can be performed in such a way that the subgraph traversed by packets from a set of sources to a given destination evolves over time. More precisely, if at time $t_1$, the subgraph $G_1$ traversed by packets originating at sources $S_1$ and $S_2$ and ending at a destination $D$ is {\it different} from the subgraph $G_2$ traversed between sources $S_1, S_2$ and destination $D$ at time $t_2$, then the intersection of $G_1$ and $G_2$ is {\it small}. So, if this subgraph evolves so that it is different at different time-slots, then for each time-slot that decoding fails (due to some node in the subgraph being malicious or faulty), the subgraph traversed during that time-slot can be isolated and intersected with subgraphs of other such time-slots (when decoding failed). This will enable the receiver to identify a small set of nodes (in the intersection) as candidates for the malfunctioning/malicious node.

The subgraph creation needs to be done carefully, so that every $k$ subgraphs (for some chosen $k$) have a nonempty but not too large intersection. We defer the details of such a construction to a future version of the paper.

\section{Numerical Results}\label{sec:simulation}
\begin{figure}
\begin{center}
\includegraphics[scale=0.5]{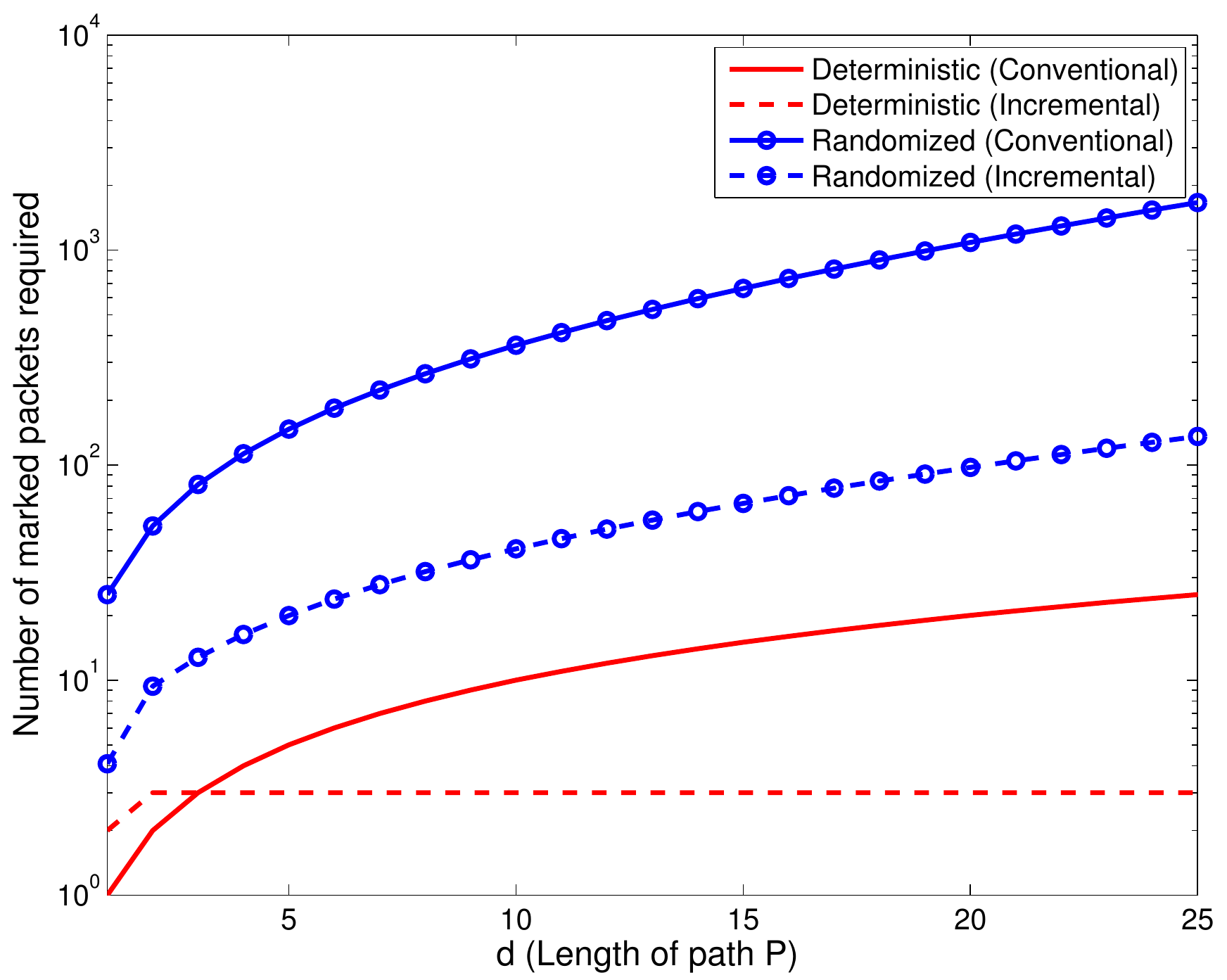}
\end{center}
\caption{Comparison of the number of marked packets needed for determining $\mathcal P$ for both deterministic and randomized full path encoding versions.}\label{fig:det}
\end{figure}

In this section, we present some numerical results on the number of market packets required to successfully perform algebraic traceback. We consider a network where the nodes have $16$-bit long IDs. This means the order $p$ of the prime field, where the identities come from, should be greater than $2^{16}-1$. We assume $p=2^{16}+1$, which is the smallest prime greater than $2^{16}-1$. Then for deterministic path encoding, for a dynamic path $\mathcal P$ of length $d$ the number of marked packets needed for determining the path initially is $d$. As derived in Section \ref{sec:fullpath}, the number of marked packets needed for determining the change in path $\mathcal P$, once its topology is known, is given by $l=\lceil\frac{\log_2 d}{\log_2 p}+\delta\rceil$, where $\delta \in {\mathbb N}$ is a constant which determines the rate with which the (union) upper-bound of the probability of error decays with $p$. We choose $\delta=2$, which upper-bounds the probability of error by $\frac{1}{p^2}$, which is approximately $2^{-32}$ for our case. Figure \ref{fig:det} makes the comparison between the number of marked packets needed for the usual non-incremental traceback and the incremental version for deterministic path encoding. As observed, the incremental version of traceback proves to be better - the number of marked packets is far smaller and the rate of growth of marked packets needed, with $d$ increasing, is also smaller than non-incremental traceback.

The average number of marked packets needed for randomized path encoding for both the non-incremental and incremental traceback versions is also shown in Figure \ref{fig:det}. Here, we consider the case when the nodes mark packets independently of each other with probability $q=0.04\,(q_i=q\,\,\forall i)$. This gives $f_0=(1-q)^d=(0.96)^d$ and $f_1=q(1-q)^{d-1}=0.04(0.96)^{d-1}$. The average number of marked packets needed by the conventional traceback is $d\left(\frac{1-f_0}{f_1}\right)$ and the average number of packets needed by the incremental traceback is $\lceil\frac{\log_2 d}{\log_2 p}+2\rceil \left(\frac{1-F_0}{F_1}\right)$. In this case, the average number of marked packets needed for incremental traceback increases significantly compared to the deterministic path encoding case, but it is still less than the number needed by conventional randomized path encoding version of traceback.

\begin{figure}
\begin{center}
\includegraphics[scale=0.5]{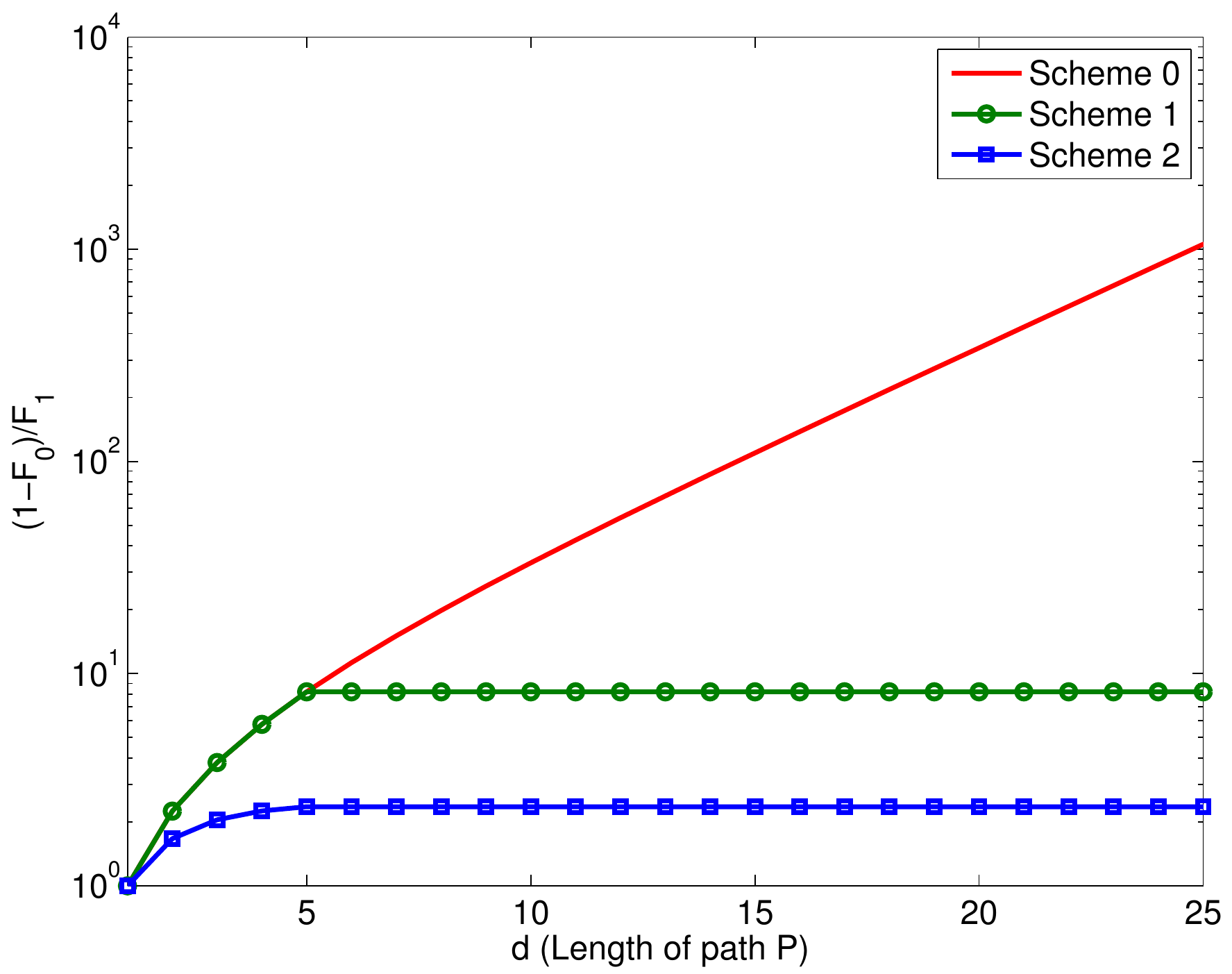}
\end{center}
\caption{Comparison of the quantity $(1-F_0)/F_1$ and its variation with respect to $d$ for various marking schemes.}\label{fig:marked}
\end{figure}

We next analyze the performances of Schemes 1 and 2 (Section \ref{subsec:marked}) in reducing the average order of marked packets needed and compare it with the scheme in \cite{Algebraic} i.e., where all nodes mark packets with same probability (let us call this Scheme 0). For both the Schemes 1 and 2, we assume $h_0=5$ i.e., once a node sees a marked packet of hop-count $5$ or more, it does not mark it. We consider $q=0.2$ for Scheme 0 and 1, $\alpha=0.5$ for Scheme 2. Then for $d\geq h_0=5$, the fraction of unmarked packets are $32\%$ and $30\%$ for Schemes 1 and 2 respectively, which seems reasonable. Figure \ref{fig:marked} depicts the variation of $\frac{1-F_0}{F_1}$ with $d$. Clearly for Schemes 1 and 2, the value becomes a constant while for Scheme 0, it continues to grow in value. Thus, Schemes 1 and 2  reduce the average order of number of marked packets needed to perform traceback.

\section{Conclusion and Remarks}\label{sec:conclude}
In this paper, we present a mechanism of performing incremental algebraic traceback in networks with a topology that is changing much slower than its rate of communication. We initialize the system using an established algebraic traceback mechanism, and then track the network as it evolves using an efficient incremental traceback mechanism. The decoding process is altered from a traditional traceback scheme. This decoding mechanism actively searches for a change in network topology in the incoming packets, and when one is detected, it determines what the change is (insertion or deletion), where it has occurred in the network and what the new ID, if any, of the inserted node is. We also show that, for the case with no ID spoofing among nodes, the resulting algorithm requires $O(\log d)$ marked packets and a complexity of $O(d\log d)$ before it can declare success in determining the ID of the change in a path of $d$ nodes. We also show, very straightforwardly, that this packet overhead is order-wise optimal.

Note that our proof mechanisms closely resemble random coding proofs in information theory for discrete additive memoryless channels. \emph{Algorithms I} through \emph{III} can be viewed as ``achievability" proofs from conventional information theory, while, in this case, the converse is straightforward. A final remark is that, when we swap a more stringent probability 1 (zero error) requirement for tracking the changing path in a dynamic network with a arbitrarily small error constraint, the resulting time taken and complexity of the incremental traceback algorithm decreases substantially.

\end{document}